# Mesoscopic Characterization of Bubble Dynamics in Flow Boilling following A Pseudopotential-based Approach

Aritra Mukherjee, Dipankar N. Basu, Pranab K. mondal

*Department of Mechanical Engineering, Indian Institute of Technology Guwahati*

**Abstract**

Present study explores the capability of the pseudopotential-based thermal lattice Boltzmann (LB) model in emulating the underlying thermohydrodynamics of flow boiling in a narrow fluidic channel. In contrary to the conventional Eulerian-averaging-based approach, it adheres to the mesoscopic Boltzmann statistical averaging, which allows natural phase separation and no need of assuming the initial interface. A narrow fluidic channel, with specified inlet temperature and flow rate, and exit pressure, housing a microheater at the bottom wall is considered as the computational domain of interest. Adopted boundary conditions ensures subcooled flow boiling through the channel, and the present algorithm successfully emulates the corresponding characteristics. The complete dynamics of bubble ebullition at the nucleation site, and subsequent flow regimes are adequately reproduced. Both bubbly and slug flow patterns are illustrated through the temporal evolution of the interface, and associated pressure drop and heat transport characteristics. Dependence of the departure characteristics on the flow rate, wall superheat and surface wettability is found to be consistent with available literature, which substantiates the competence of the present algorithm.

## 1. Introduction

Precise numerical simulation of multiphase flow is one of the most arduous tasks in purview of computational fluid dynamics, the primary contributor towards such intricacy being the discontinuous property variation across a dynamic and malleable interface of infinitesimal thickness. The scenario is particularly delicate with liquid-vapor mixtures, owing to the involvement of high density and viscosity ratios. Such flow situations, however, are very much ubiquitous both in nature and industrial processes, encompassing large-scale power boilers, cooling towers and airconditioners, to the microarteries of cardiovascular system. Rapidly-increasing focus on miniaturization and effort to mimic biological systems in heat transport devices have projected flow boiling through narrow fluidic channel as a very lucrative application, particularly with its potential of offering high heat flux with large areato-volume ratio and small temperature differential. That has envisioned a new paradigm of experimental and numerical research with multiphase thermohydrodynamics in the present millennium, a comprehensive perspective of which is available in Kandlikar[1]. Possible involvement of multiple length and time scales, dependence of the phenomenon on the distribution of nucleation sites, enhanced role of surface topology and wettability, and lesser reliance on gravity are some of the additional factors that can have pivotal influence in mini- and microscales, thereby demanding dedicated approach for appraisal. Incompetence of common measuring tools at smaller scales and hindrance in optical assessment due to refractive interface imposes enhanced reliance on numerical procedures, stimulating the development of several relevant techniques[2,3].

One major challenge in multiphase modeling is the identification of suitable form of macroscopic governing equations. For a boiling channel experiencing flow of liquidvapor mixture, both phases can notionally be regarded as distinct fluids, characterized by individual velocity and temperature fields, and separated by multiple, deformable and moving interfaces, with mass, momentum and energy interactions across the same. Realization of a phase as continuous or dispersed differs widely with the flow regimes[4,5], proclaiming for the adoption of some averaging procedure to derive effective conservation equations and incorporation of appropriate, often empirical, closure relations. A multitude of averaging approaches have been proposed over the years[6-9], with one common frailty being the uncertainty in exactly locating a constituent at a particular instant, consequently limiting the predictions mostly to gross thermalhydraulic characterization. On the contrary, the local instant formulation[10], despite theoretically being an excellent option for separated flows, involves added mathematical difficulty and greater computational resource requirement, while also necessitating the validity of continuum within each subregion. That, however, can act as the base for developing the macroscopic conservation equations following suitable averaging approaches. The Eulerian-Lagrangian description is especially amenable to particulate flows[11-14], but is commonly not favored for diffusion-dominated or phase-change problems. The Eulerian-Eulerian averaging is more popular, owing to its direct relevance to human observation in terms of the time-space description of the physical phenomenon, yielding numerous relevant research efforts. Conventional multiphase computation, however, requires a precise algorithm for coupling the interfacial information with the conservation equations, which can either be a surface-based interface-tracking or volume-based interface-capturing approach. While the classical MAC/SMAC/SMMC-based algorithms[15,16] belong to the former category, techniques like level-set[17,18], volumeof-fluid[19,20] and phase-field[21] are popular examples of the later. Several logical combination of these methods are also available in literature[22-24], allowing a gradual development in associated concepts. Quite a few recent efforts to apply similar methodology for analyzing boiling in smaller dimensions are also available[25-31]. However, as was observed by Kharangate and Mudawar[32], most of the reported multiphase computational efforts are restricted either to pool boiling or simplified flow boiling, consequently limiting themselves mostly to validation studies. A major impediment for Euler-averaged algorithms is the requirement of pre-defined vapor embryo, to initialize the interface, which makes them vulnerable for realistic flow boiling situations. A possible remedy to the above conundrum can be contrived through the Boltzmann statistical averaging, which uses the concept of particle number density, and emphasizes on the collective particle mechanics with increase in the number of particles within the domain and their respective interactions. It is well-established that such approach involving large number of particles with an appropriate choice of meanfree-path can efficiently emulate the continuum mechanics[10]. Such a mesoscopic perspective, while being more fundamental, offers easy implementation of boundary conditions and fully-parallel algorithm. Consequently the lattice-Boltzmann method (LBM)[33,34] has emerged as a strong contender for large-scale multiphase simulation over last couple of decades, with its linear convection operator in velocity-space and second-order numerical accuracy in space and time through multiscale expansion being particularly alluring. The discreteparticle-based approach allows natural phase separation, assuaging the need of any interface tracking or empirical relations. Following the pioneering attempt of Rothman and Keller[35], several multiphase-LB algorithms have been proposed, and the pseudopotential-based approach, originally formulated by Shan and Chen[36,37], has found particular favour from the boiling community, as can be substantiated through the large volume of available literature[38-45]. It employs a nearest-neighbor interaction model, which is a close approximate to the Lennard-Jones potential, allowing efficient computation and relatively-smooth representation of the interface. Despite reasonable success with pool boiling simulations, application of Shan-Chen-LB model (SC-LBM) for flow boiling is quite scarce till date. Gong and Cheng[46] were probably the first ones to simulate saturated flow boiling at low Reynolds number through a horizontal microchannel at the absence of gravity, extending their dual-distribution-LB methodology[40]. Slug flow regime and logical parametric effects were observed, despite unrealistic pressure variation across the interface. Sun et al.[47] employed the same philosophy in a vertical channel of 7.5 aspect ratio, designed with several discrete nucleation sites on both walls, with subcooled inlet temperature and were able to reproduce the associated physics. The effect of contact angle was considered in the follow-up study[48]. Above three studies are testimony towards suitability of SC-LBM for flow boiling and also bestow motivation for the present work, as we look to fill the void in open literature by performing LBM simulation of subcooled flow boiling through a horizontal narrow fluidic channel at the presence of gravity. Primary focus is on characterizing the flow regimes in terms of the departure diameter and frequency as functions of input parameters, such as, inlet mass flux, wall superheat and surface wettability. Conscientious effort is made to ensure the replication of the underlying microdynamics and also the development of sample flow-regime map at small dimensions.

A brief rundown of the manuscript organization is mentioned here. Section-II details the mathematical formulation of the pseudopotential-based LB algorithm, whereas the results obtained from the in-house code are discussed in Section-III. Finally, main conclusions, along with the possible directions of future research, are summarized in Section-IV.

## 2. Mathematical Formulation

The governing equation in LB philosophy is framed in terms of the particle distribution function (PDF) $f(\bar{x},\bar{\varsigma},t)$ which symbolizes local instantaneous population density at lattice-level. Here we adhere to the double-distributionfunction model of Gong and Cheng[40], introducing a second distribution function $g(\bar{x},\bar{\varsigma},t)$ to represent the energy distribution in terms of the local lattice-scale temperature.

### A. Pseudopotential based LB model

The PDFs are probabilistic density distribution functions, which signifies the mass density at both physical- and latticespace. Their evolution in the $i^{th}$ lattice direction can be presented as,

$$f_i(\bar{x}+\bar{c}_i\Delta t, t+\Delta t) = f_i(\bar{x},t) + \Omega_i(\bar{x},t) + \bar{F}_i(\bar{x},t) \quad (1)$$

where $f_i$ is the PDF with velocity $\bar{c}_i$ at position $\bar{x}$ at time $t$. $\Omega_i$ corresponds to the inter-particle collision operator, and is popularly represented following the BGK model[34], which allows the population to relax towards the equilibrium distribution $\left(f_i^{eq}\right)$ based on the relaxation time $\left(\tau_f\right)$ as :

$$\Omega_i\left(\bar{x},t\right) = -\frac{\Delta t}{\tau_f}\left[f_i\left(\bar{x},t\right) - f_i^{eq}\left(\rho,\bar{u}\right)\right] \quad (2)$$

We adopt the following form of the equilibrium distribution function,

$$f_i^{eq}\left(\rho,\bar{u}\right) = w_i \rho \left[1 + \frac{\bar{c}_i \cdot \bar{u}}{c_s^2} + \frac{1}{2}\frac{\left(\bar{c}_i \cdot \bar{u}\right)^2}{c_s^4} - \frac{1}{2}\frac{\bar{u}^2}{c_s^2}\right] \quad (3)$$

whereas the relaxation time is governed by the kinematic viscosity of the concerned fluid as $\upsilon = c_s^2\left(\tau - \frac{1}{2}\right)\frac{\Delta x^2}{\Delta t}$. The details of the imposed weight factor $w_i$ for the selected D2Q9 laattice configuration is available in Fig. 1, which also enumerates the relevant velocity vectors $c_i$. Here $\bar{c}_s = \bar{c}/\sqrt{3}$ defines the acoustic speed of the lattice, where $\bar{c} = \Delta\bar{x}/\Delta t$ is the lattice velocity, and $\Delta x$ and $\Delta t$ are the grid spacing and time step respectively.

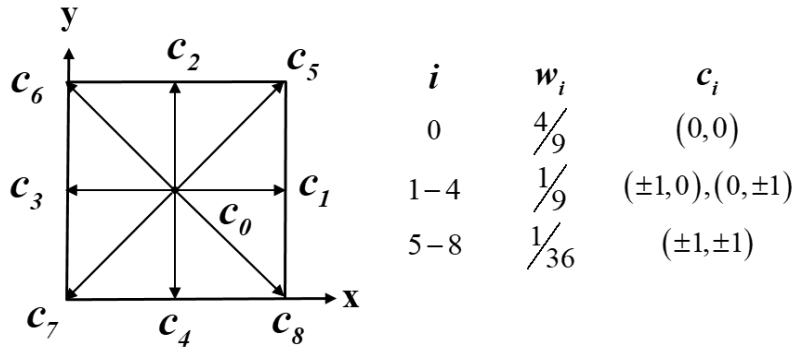

**FIG. 1**. Vectorial representation of the D2Q9 lattice adopted in the present study; here $c_i$ is the unit vector in the $i^{th}$-velocity direction and $w_i$ is the corresponding weight factor.

The lattice-level mass density can be computed by taking the zeroth-order velocity moment of the PDF as,

$$\rho\left(\bar{x},t\right) = \sum f_i\left(\bar{x},t\right) \quad (4)$$

The term $\bar{F}_i\left(\bar{x},t\right)$ appearing in Eq. 1 corresponds to the prevailing body force field, and that is estimated following the exact difference method[49] shown below.

$$\bar{F}_i(\bar{x},t) = f_i^{eq}\left(\rho(\bar{x},t), \bar{u}^{eq} + \Delta\bar{u}\right) - f_i^{eq}\left(\rho(\bar{x},t), \bar{u}^{eq}\right) \tag{5}$$

where $\Delta\bar{u} = \dfrac{\bar{F}\Delta t}{\rho}$ and $\bar{u}^{eq}$ is the equilibrium velocity. The equilibrium velocity is written as:

$$\bar{u}^{eq}(\bar{x},t) = \sum f_i(\bar{x},t)\bar{c}_i \tag{6}$$

The actual fluid velocity $(\bar{u}^{act})$ is different from $\bar{u}^{eq}$ and calculated as:

$$\bar{u}^{act}(\bar{x},t) = \bar{u}^{eq}(\bar{x},t) + \dfrac{\bar{F}\tau_f \Delta t}{2\rho} \tag{7}$$

The body force $\bar{F}$ has contribution from the inter-particle interaction, gravity and surface wettability. The inter-particle interaction force, popularly referred as the Shan-Chen force $\left(\bar{F}^{SC}\right)$ is the one responsible for the natural phase separation at the mesoscopic level. It can be estimated as[50],

$$\bar{F}^{SC}(\bar{x}) = -\beta c_0 g \psi(\bar{x})\nabla\psi(\bar{x}) - (1-\beta)c_0 g \nabla\psi^2(\bar{x})/2 \tag{8}$$

Here $\beta$ is a weighting factor associated with the equation of state, which is set as 1.16 following Gong and Cheng[50]. The discretized form of the above equation, calculating the interaction force between particles at the lattice site $\bar{x}$ and its nearest neighboring site $\bar{x}_1$, can be written as,

$$\bar{F}^{SC}(\bar{x}) = -\beta\psi(\bar{x})\sum_{\bar{x}_1} G(\bar{x},\bar{x}_1)\psi(\bar{x})\cdot(\bar{x}_1 - \bar{x}) - \dfrac{(1-\beta)}{2}\sum_{\bar{x}_1} G(\bar{x},\bar{x}_1)\psi^2(\bar{x})\cdot(\bar{x}_1 - \bar{x}) \tag{9}$$

where $G(\bar{x},\bar{x}_1)$ represents the interaction strength between the particles. Value of $G(\bar{x},\bar{x}_1)$ is given as:

$$G(\bar{x},\bar{x}_1) = \begin{cases} g_1 & \text{if } |(\bar{x}-\bar{x}_1)| = 1 \\ g_2 & \text{if } |(\bar{x}-\bar{x}_1)| = \sqrt{2} \\ 0 & \text{otherwise} \end{cases} \tag{10}$$

with $g_1 = 4g_2$ following the work of Yuan et al.[51].

Here $\psi(\bar{x})$ is the pseudopotential which depends on the local density of a particular lattice node and forms the basic structure of the interparticle interaction force. It is written as:

$$\psi(\rho) = \sqrt{\frac{2(p - c_s^2 \rho)}{c_0 g}} \quad (11)$$

where $c_0 = 6.0$.

The physical requirement of having coexistence of phases or components puts a constraint on the choice of equationof-state (EOS), which describes the complex interdependency among pressure, density and temperature. Two phases of a fluid can coexist at specified densities at a specific temperature and those densities are determined by the coexistence curve obtained from Maxwell area construction rule[34]. In the present model, no specific temperature can be defined, and hence the coexisting densities cannot directly be obtained[52], which can lead to a possible thermodynamic inconsistency with the original SC scheme. However, with the modifications proposed in recent years[50,51], there is marked improvement in thermodynamic consistency and the same is followed here as well. We replace pressure (p) in the Eq. 11 by the Peng-Robinson EOS (PR-EOS), which adopts the following form at the lattice-level.

$$p = \frac{\rho R T}{1 - b\rho} - \frac{a\alpha(T)\rho^2}{(1 + 2b\rho - b^2\rho^2)} \quad (12)$$

where $\alpha(T) = \left[1 + \left(0.37464 + 1.54226\omega - 0.26992\omega^2\right) \times \left(1 - \sqrt{T/T_c^2}\right)\right]^2$

$a = 0.45724 R^2 T_c^2 / p_c$ and $b = 0.0778 R T_c / p_c$

Here $\omega$ is a fluid-dependent ascentric factor. We use R134a as the working fluid for all the simulations reported in the present work, for which $\omega = 0.32$. Following Yuan et al.[51], the values of a and b are set as 2 / 49 and 2 / 21 respectively, by scaling them using the universal gas constant R, which is assigned with a lattice-level value of unity. The principle of corresponding states is enforced to relate the physical and lattice-level quantities, as the reduced properties must remain identical at all scales.

$$\rho_R = \frac{\rho^{LB}}{\rho_c^{LB}} = \frac{\rho^{real}}{\rho_c^{real}}, \quad T_R = \frac{T^{LB}}{T_c^{LB}} = \frac{T^{real}}{T_c^{real}}, \quad p_R = \frac{p^{LB}}{p_c^{LB}} = \frac{p^{real}}{p_c^{real}} \quad (13)$$

The buoyancy force is modeled following Kang et al.[53] as,

$$\bar{F}^g(\bar{x}) = \bar{g}_a \times \left(1 - \frac{\rho_{avg}}{\rho_i}\right) \quad (14)$$

where $\rho_{avg}$ is the average density of the whole computational domain, $\rho_i$ is the local density and $\bar{g}_a$ is the gravitational acceleration.

Finally the surface wettability force $\bar{F}^{wet}$ is incorporated following a very simple model proposed by Benzi et al.[54] for a single-component multiphase flow and solid wall. Here the desired contact angle is established by tuning a parameter $\rho_w$, identified as the false wall density. It can be viewed as the density at the solid node in immediate vicinity to the interface. Consequently, the force of adhesion at the fluid-solid interface is calculated as,

$$\bar{F}^{wet}(\bar{x}) = -G_{wet}\psi(\bar{x})\sum w_i \psi(\rho_w) S_{ind}(\bar{x}+\bar{c}_i\Delta t)\bar{c}_i\Delta t \tag{15}$$

An indicator function $S_{ind}$ as appearing in Eq. (15) is used to denote the solid and fluid nodes. Note that the indicator function takes a value 1(one) for a solid node to calculate the adhesive force. On the other hand, for a fluid node this function becomes 0 (zero).

## B. Energy Transport equation

A second distribution function is defined for describing the thermal field following Gong and Cheng[40], and its evolution can be tracked using the following energy equation.

$$g_i(\bar{x}+\bar{c}_i\Delta t, t+\Delta t) = g_i(\bar{x},t) + \frac{g_i - g_i^{eq}}{\tau_g} + w_i \Delta t \phi_i \tag{16}$$

where $\phi$ is the source term and used for the latent heat estimation and the relaxation parameter is a direct function of the thermal diffusivity as $\alpha_T = c_s^2\left(\tau_g - \frac{1}{2}\right)\frac{\Delta x^2}{\Delta t}$. The lattice-level temperature can be computed using the moment of this distribution function as,

$$T(\bar{x},t) = \sum g_i(\bar{x},t) \tag{17}$$

and the equilibrium temperature distribution function can be expressed as,

$$g_i^{eq}(\rho,\bar{u},t) = w_i T(\bar{x},t)\left[1 + \frac{\bar{c}_i \cdot \bar{u}^{act}}{c_s^2} + \frac{1}{2}\frac{(\bar{c}_i \cdot \bar{u}^{act})^2}{c_s^4} - \frac{1}{2}\frac{(\bar{u}^{act})^2}{c_s^2}\right] \tag{18}$$

Choice of energy source term is very crucial from the perspective of numerical accuracy. Házi and Márkus[55] proposed the following form by introducing the thermodynamics *Tds* relations into the macroscopic energy equation.

$$\phi = \frac{T}{\rho^2 c_v}\left(\frac{\partial p}{\partial T}\right)_\rho \frac{d\rho}{dt} + T\nabla \cdot \bar{u} \qquad (19)$$

where $\left(\frac{\partial p}{\partial T}\right)_\rho$ is calculated directly from the corresponding EOS. It may be mentioned here that the numerical estimation of the temporal density gradient can be quite inconvenient. Therefore, an alternate version of Eq. 19 was suggested by Gong and Cheng[40] by introducing the continuity equation into it, as shown below.

$$\phi = T\left[1 - \frac{1}{\rho c_v}\left(\frac{\partial p}{\partial T}\right)_\rho\right]\nabla \cdot \bar{u} \qquad (19)$$

This particular form was found to be easier for numerical implementation and computationally less-expensive, and hence adhered with in our work. Second order central difference scheme is used to discretize the divergence of velocity and temporal gradient of density. A flowchart for the complete algorithm is presented in Fig 2.

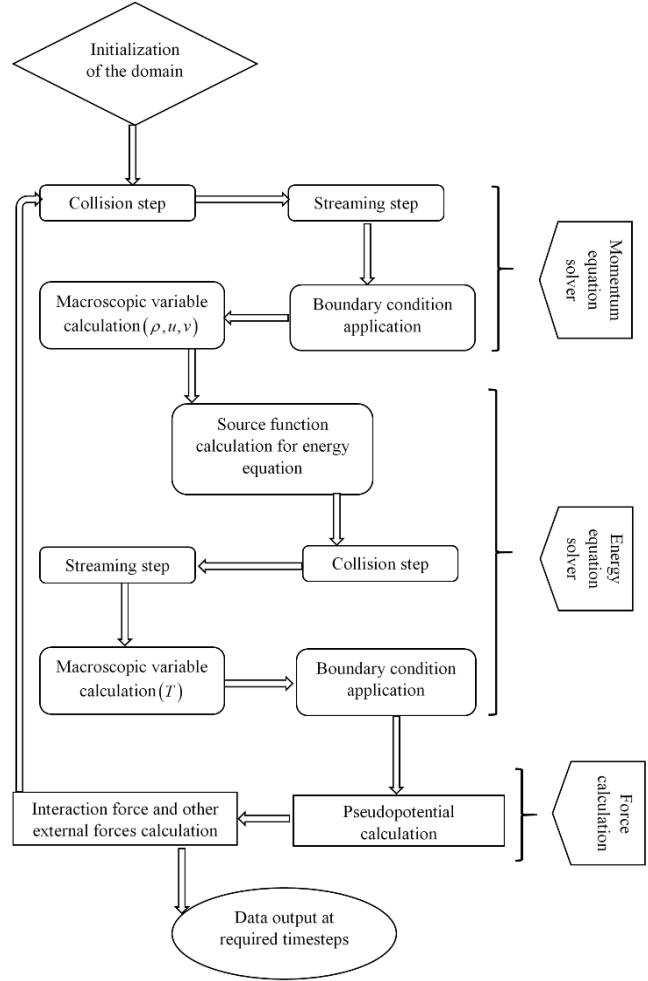

**FIG. 2.** A flowchart of the complete algorithm, encompassing the pseudopotential-based LB model and the energy equation

## C. Problem Definition

A computational domain comprising of $40 \times 1000$ lattices is selected to represented a two-dimensional rectangular channel having height-to-length ratio of $1/25$, the schematic of which is presented in Fig. 3. The channel is subjected to a stream of subcooled liquid with specified velocity and temperature at the inlet, whereas constant pressure and zero temperature gradient conditions are imposed at the exit plane. Both the top and bottom walls are stationary impermeable surfaces, allowing for no-slip boundary condition, while remaining isothermal as well. A microheater is mounted on the bottom wall at one-fifth distance from the channel inlet $(L_h)$. It is numerically replicated by inflicting a constant degree of wall superheat on five lattice nodes. All the boundary conditions are marked in Fig. 3 for easier cognizance. The no-slip boundary condition is numerically realized through the classical bounceback approach[34], whereas the entry and exit conditions are accomplished using the non-equilibrium bounceback scheme[56]. The anti-bounceback approach[34] is preferred for the thermal boundary conditions.

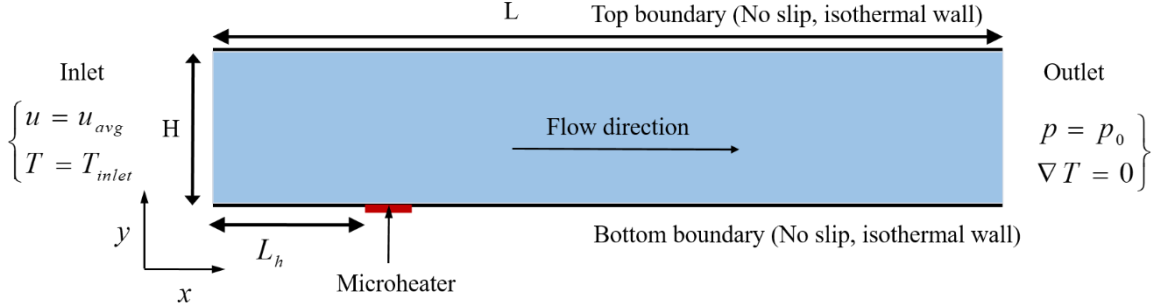

**FIG. 3.** (Color online) Schematic presentation of the physical domain considered in present study, along with all applicable boundary conditions. The coordinate system is attached at the left bottom corner of the channel.

The domain is initially assumed to be filled with saturated liquid maintaining its temperature at $T_R = 0.9$. As mentioned earlier, R134a is considered to be the working fluid and all thermophysical properties are estimated at this temperature. Computation is carried out without heating for 10000 timesteps since triggering the motion, to allow the flow be hydrodynamically fully-developed. Only then the microheater is switched on and ensuing hydrodynamics are followed for computational characterization.

### D. Model validation

In an effort to ascertain the accuracy of the algorithm, certain benchmark cases from literature are simulated and compared with the standard solutions. Our first selection is for single-phase isothermal flow through a duct, primarily to check the consistency of the solver. The cross-sectional profile of x-direction velocity is compared with the analytical solution in Fig. 4(a), for the axial position of 8H with $\text{Re}_{in} = 0.96$ at t = 10000 lattice units. As desired, an excellent match can be observed, with parabolic shape and maximum velocity assuming 1.5 times the average value, which substantiates the correctness of the algorithm for single-phase flows.

As noted earlier, one major concern with classical SC model is the thermodynamic inconsistency, as the simultaneous requirement of compliance to EOS and consistency to the thermodynamic definition of surface tension can be satisfied only for very low density ratio. The recent amendments[50,51], though, has helped eliminating this apprehension till certain range. According to the Maxwell area construction rule, for a given temperature, liquid-vapor coexistence is possible only for a specific pressure, characterized by two specific density values. To validate the same, the standard static bubble test is performed, where a vapor bubble is placed in an infinite pool of liquid, simulated as a periodic domain. Regardless of the initial values, both the phases gradually align their respective densities solely as a function of the specified saturation temperature and the bubble attains a stable radius, after certain timesteps. Reduced density values $(\rho_R)$ are plotted against the reduced temperature $(T_R)$ in Fig. 4(b), to facilitate a direct comparison with the analytical value. Satisfactory level of conformity is obtained for $T_R \geq 0.7$. In the present study, we persist with $T_R = 0.9$, where the model is very much consistent thermodynamically, and hence is expected to yield accurate predictions.

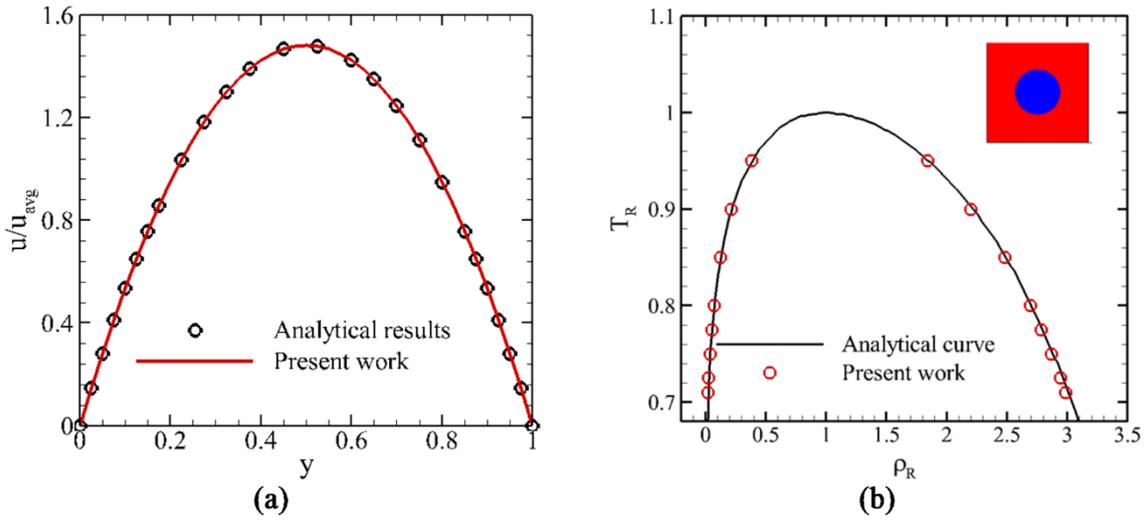

**FIG. 4.** (Color online) Model validation: (a) Parabolic velocity profile for single-phase flow through rectangular dust shows excellent match with the analytical solution; (b) Coexisting densities of liquid and vapor phases adhere to the Maxwell area construction rule till $T_R = 0.7$

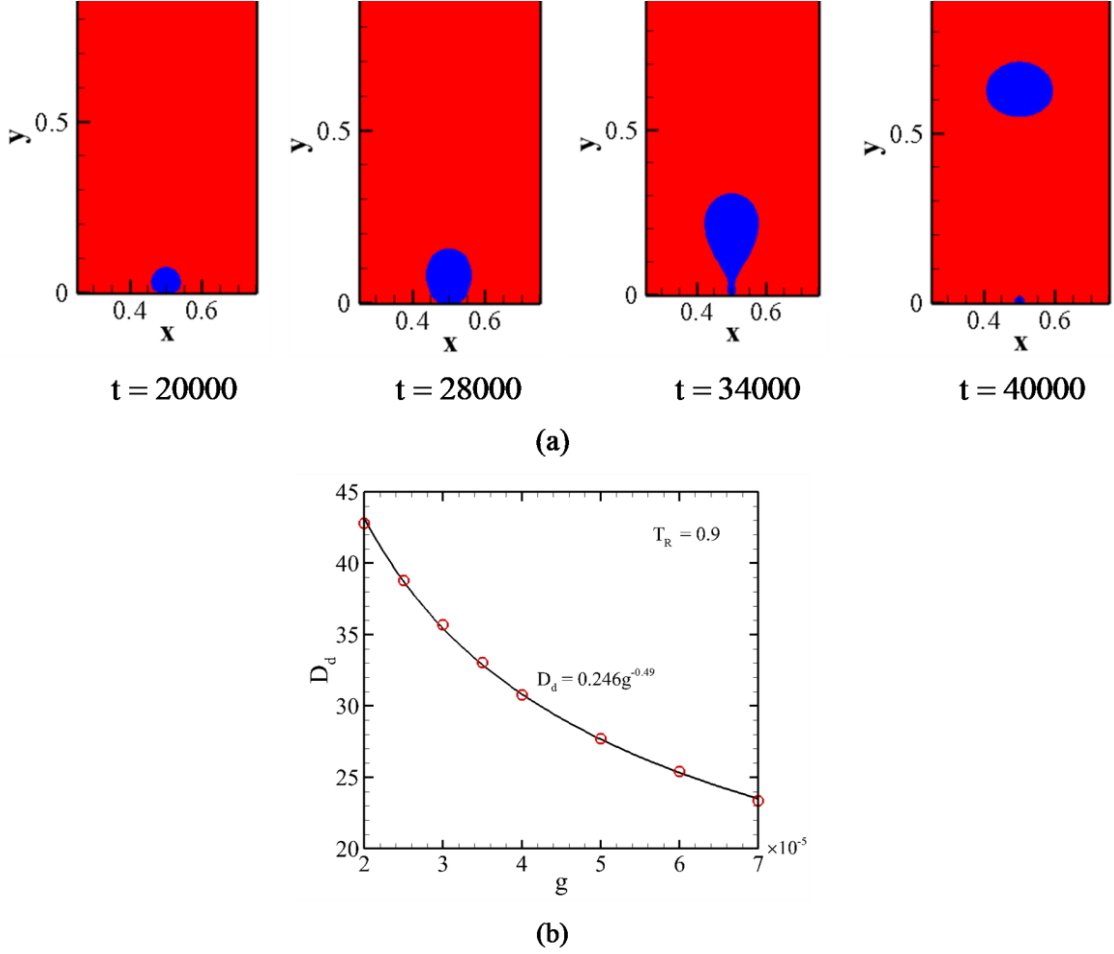

**FIG. 5.** (Color online) (a) Snapshots of one bubble ebullition cycle from a microheater in an open vertical domain at different time instants, with red and blue colors respectively symbolizing liquid and vapor phases; (b) Variation in bubble departure diameter with gravitational acceleration shows reasonable allegiance to the Fritz's correlation[57]

Owing to the lack of reliable experimental data, as well as thermalhydraulic correlation, for flow boiling at this temperature range, another set of validation is attempted for pool boiling scenario. According to Fritz's correlation[57], the bubble departure diameter is reliant on gravity, surface tension, surface wettability, as well as the coexisting densities of both phases. For a specified temperature, other properties are constant, making the departure diameter a direct function of gravity. We consider a rectangular domain of aspect ratio 4, initially filled with saturated liquid having $T_R = 0.9$. The bottom boundary is maintained isothermal at the saturation temperature itself, while the side boundaries are periodic and the top is envisaged as a free surface. A microheater spanning 5 lattice units is placed at the center of the bottom wall, which is maintained at a higher temperature of $T_R = 1.05$. Snapshots of the resultant ebullition cycle is presented in Fig. 5(a), vividly illustrating each important stage of pool boiling, such as nucleation, growth and departure, for a lattice-level gravitational acceleration of $g = 5 \times 10^{-5}$. The variation in the departure diameter for different levels of g is portrayed in Fig. 5(b) for the same temperature combinations mentioned above. The developed data points can be regressed as, $D_d = 0.246 g^{-0.49}$ which is within acceptable proximity to the exponent of 0.5 proposed by Fritz. Therefore, we can definitely claim that the present SC-LBM algorithm can successfully reproduce boiling phenomenon and hence can be used to explore the microdynamics of flow boiling.

## 3. Results and discussion

It has already been highlighted that the objective of the present study is to explore the dynamics of vapor bubbles during flow boiling in a narrow fluidic channel. During real-life experimentation, vapor production is facilitated through the presence of nucleation sites, in the form of cavities or discontinuities, on the heated surface. It is also possible to embed a microheater, having dimension comparable to the bubble diameter, on the wall adjacent to the flow field, which can serve as a nucleation site. A direct numerical duplication of the same is not possible with the Eulerian-averaging approach. It is necessary either to initialize the domain with a pre-existing vapor nucleus[26], or allow vapor injection through a microgap till the formation of a stable nucleus, which is referred as pseudo-nucleation[28]. Both the options fail to mimic the experimental situation, and phase-change LB models clearly score better here by allowing natural separation of phases. In the present SC-LBM algorithm, liquid-vapor coexistence density is controlled by the selected non-ideal EOS (PR-EOS) appearing inside the modified pressure tensor. Once the liquid temperature is higher than saturation (superheated liquid) and sufficient energy is available in latent mode, the EOS will allow the nodal instantaneous density to acquire the magnitude corresponding to saturated vapor. The reverse is true on dissipation of sufficient amount of energy from the vapor phase. It is, therefore, possible to initiate vapor nucleation in the flow field by placing a microheater on the surface (Fig. 3), analogous to the experiments, consequently allowing us to get a more comprehensive perspective of flow boiling. We, therefore, employ the SC-LBM algorithm for exploring the dynamics and regimes flow boiling in the subsequent sections. Unless mentioned otherwise, all the reported simulations correspond to $T_{R,in} = 0.9 T_c$, $\text{Re}_{in} = 0.96$, $T_{R,h} = 1.2 T_c$ and $\theta = 52.23°$.

## A. General bubble dynamics

The microheater is activated at t = 10000, with the logical presumption of the single-phase fluid already attaining fullydeveloped condition. Because of the external heating, temperature of fluid in the vicinity of the heater increases and vapor nucleation process is gradually initiated. It is a wellestablished fact that multiphase flow through any duct always experiences grater pressure drop compared to its single-phase counterpart, mostly owing to the emergence of strong accelerational component[4]. We have initialized the domain with liquid maintained at saturation temperature corresponding to the exit pressure, which is also the inlet value. On the incipience of flow, inlet pressure must increase to accommodate for the pressure drop suffered across the duct. Such rise in inlet pressure is even more prominent on the onset of nucleation, and can mutate with the transition in flow regimes. The direct implication of such pressure inflation, with constant inlet temperature, is the fluid attaining subcooled state at the entrance plane, and the entire thermohydrodynamics inside the channel assuming the characteristics of subcooled flow boiling.

Following the recent argument of Du et al.[58], during its growth from a nucleation site, a bubble can be subjected to six different forces, namely, buoyancy, contact pressure force, lift, drag, bubble growth force and surface tension. While the first three act solely in the vertical direction, drag is horizontal, and the other two can have components along both coordinate directions in a 2D domain. Combined effect of drag, lift and buoyancy is an attempt to detach the bubble from the nucleation site, whereas surface tension and bubble growth forces intend the opposite. Bubble departure is possible when the balance within these two conflicting groups is broken, and subsequently the bubble may briefly slide along the channel wall depending on the relative strength of the first set. Of course, the bubble will finally encroach into the bulk stream to flow downstream and possibly rise towards the upper part of the channel owing to buoyancy.

The rate of bubble growth is modulated by the interplay between the rate of evaporation, at the interface in contact with the slender superheated liquid layer engulfing the microheater surface, and the rate of condensation, at the portion of the interface far away from the heated surface and submerged in the subcooled liquid. During the entire ebullition cycle, encompassing nucleation, growth and departure, the rate of evaporation is much higher, as the bubble is expected to be primarily enveloped by the superheated liquid. That is, however, not true for the post-departure period, as the bubble moves with the bulk stream and is surrounded by saturated or subcooled liquid, which can lead to a reduction in bubble volume owing to the first bubble is fully condensed, second bubble has risen towards the top wall and the third bubble is about to depart.

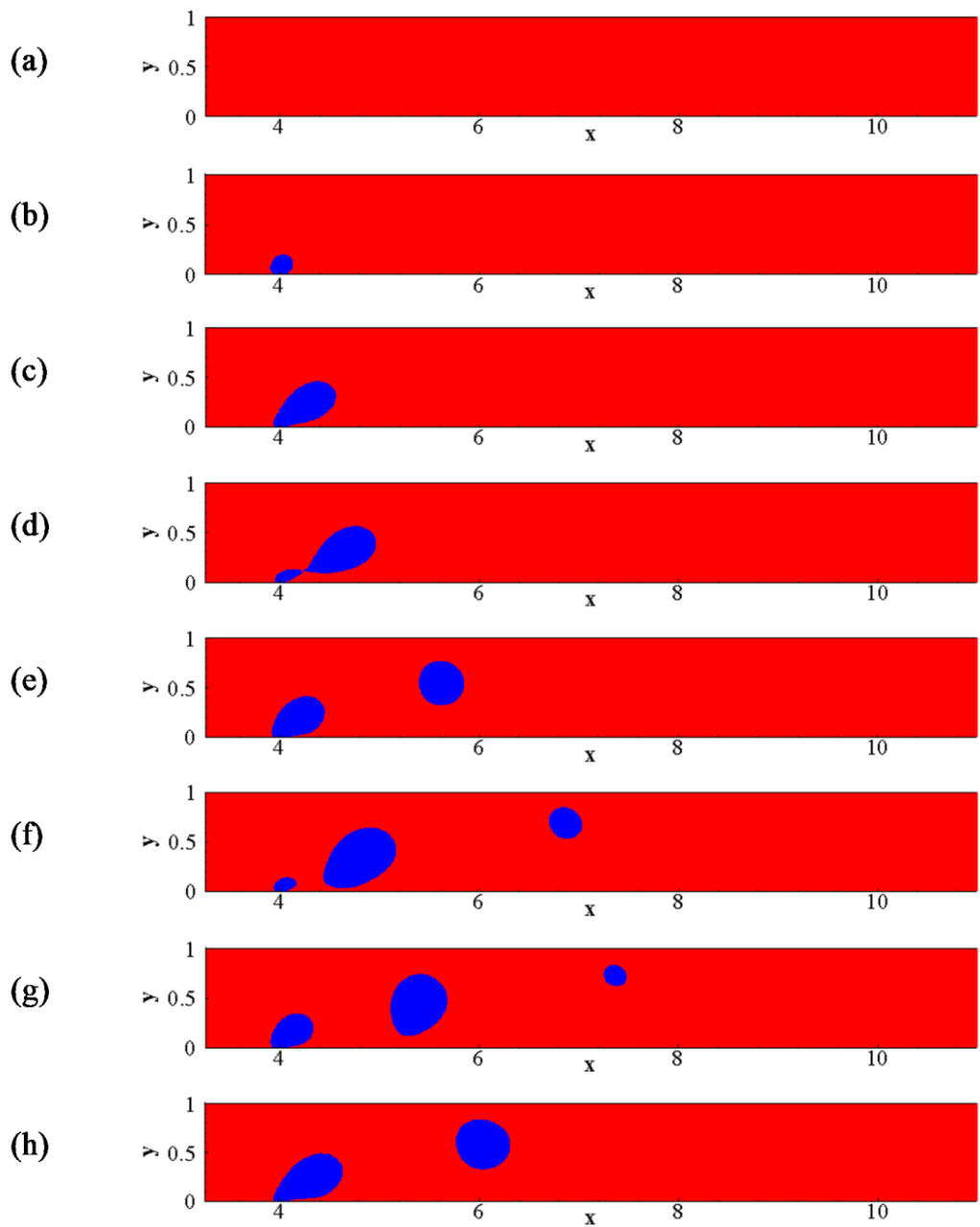

**FIG. 6.** (Color online) Snapshots of a complete bubble evolution cycle in bubbly flow ($Re_{in}$ = 0.96), with red and blue colors respectively symbolizing liquid and vapor phases; (a) t = 10000: domain filled with single-phase liquid; (b) t = 12000: appearance of small vapor embryo on the heater surface; (c) t = 20000: rapid bubble growth due to phase change; (d) t = 25848: bubble about of depart from the heated surface, with clear formation of a neck on the rear edge of the departing bubble; (e) t = 32000: first bubble is migrating as well as condensing in the bulk stream, whereas a second bubble is growing on the heater; (f) t = 38000: first bubble continuing to condense, while the second bubble has detached from the site and sliding along the bottom wall; (g) t = 42000: first bubble is about to disappear due to condensation, while second bubble has entered the bulk stream and has started to condense. Growth of the third bubble on the heater is also visible; (h) t = 47000:

dominant condensation effect and eventual collapse. The entire bubble flow dynamics explained above is demonstrated through multiple snapshots in Fig. 6 for three sequential bubble release.

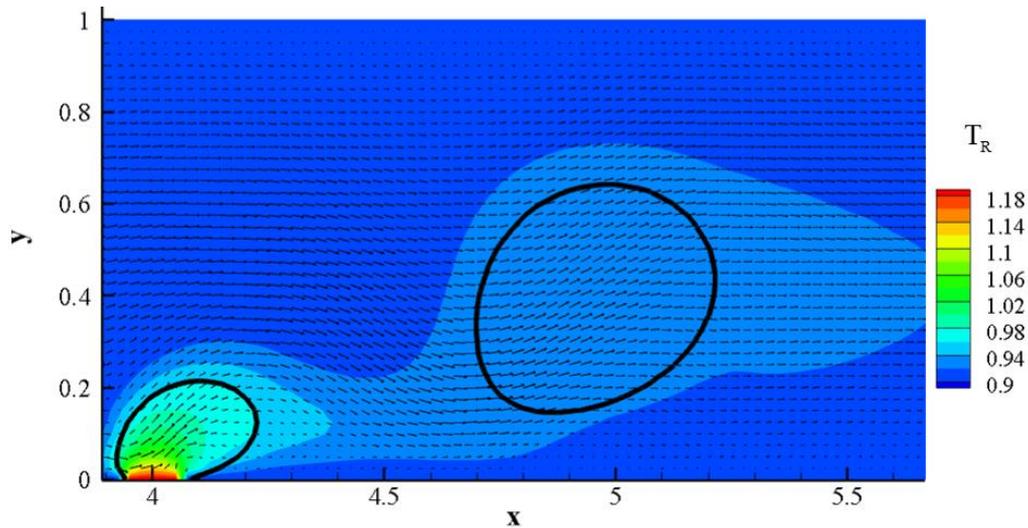

**FIG. 7.** (Color online) Contours of reduced temperature and velocity vectors around the departing bubble at t = 28000; subcooled liquid is seen to rush in to fill the void created by departure. Strong temperature gradient is also evident inside the growing nucleus.

The first instance presented in Fig. 6(a) corresponds to t = 10000, which marks the end of no-heating simulation. As expected, the entire domain is filled with single-phase liquid and the flow is hydrodynamically fully-developed. Now the microheater is activated, instigating the development of superheated liquid layer around it, and subsequent phase conversion. At t = 12000, a small vapor embryo can be spotted on the heater surface (Fig. 6(b)), which is being pushed downstream by the inertia of the flowing liquid. Consequently, the interface is asymmetric with respect to the vertical direction, with a little tilt towards right. The nucleus continually grows in volume and a much larger bubble can be seen in Fig. 6(c) at t = 20000. It is very much inclined towards right, with the principal axis making an angle of about 40 to the wall. Continuous shoving of the liquid also aids the necking of the bubble, as is hinted here, leading to the eventual departure at t = 25848 (Fig. 6(d)). As the first bubble leaves the surface, a small embryo is left behind, which starts growing into the second bubble, thereby initiating a repetitive pattern. It is interesting to observed from Fig. 6(e), while the second bubble has evolved significantly in volume, the departed one has withered a bit, suggesting towards condensation. For a better appraisal, a magnified view of the space around the microheater is presented in Fig. 7, shortly after the departure of the first bubble (t = 28000). While the second bubble sitting on the microheater is completely surrounded by superheated liquid, that is not the case for the other one. The velocity vectors clearly signify that the subcooled liquid is rushing in to fill the void created by the departure of the earlier bubble, which leads to the heat loss from the initially superheated bubble, and hence condensation. Strong temperature gradient can also be noted inside the growing nucleus, which is consistent with the experimental observations, but commonly ignored during numerical simulation by assuming the vapor phase to remain saturated.

Each of the subsequent bubbles demonstrate cycles identical to the first one, as can be affirmed following 6(f-h). Following their individual departure, the bubbles condense steadily in the subcooled stream. As mentioned before, based on the imposed inlet temperature and exit pressure, bulk liquid is mostly subcooled within the channel, while the particles interior to the interface are expected to superheated. We have already observed the same from the temperature contours presented in Fig. 7 for both the vapor bubbles. The wall superheat employed on the microheater surface $\left(\Delta T_{\text{sup}} = 0.3 T_c\right)$ possibly is not very high and hence not capable of imparting sufficient energy into the departing bubbles to make them survive in the bulk. The bubble also moves upward owing to buoyancy during its travel. As the top wall is maintained at the inlet temperature throughout, the liquid in contact with it is always highly-subcooled, and hence absorbs energy from the moving bubble, augmenting the condensation rate. The first vapor bubble completely condenses slightly before t = 47000 (Fig. 6h), when the third one is about of leave the nucleation site.

An alternate perspective to the entire lifespan can be obtained following the temporal variation in the bubble area, as illustrated in Fig. 8. As we are performing a 2D simulation, the area can be considered to be a direct representation of the bubble volume. It becomes non-zero immediately on applying wall superheat and increases monotonically till the departure. Some minor ripples can be noted immediately before departure, which is associated with the necking. On departure, the bubble leaves behind a small embryo to facilitate the next nucleation, which explains the sudden drop in area at t = 25848. As the bubble moves away from the heater, it experiences condensation owing to the heat loss to neighboring subcooled liquid. That is clearly depicted through the steady decline in the area post-departure. The rate of deterioration enhances with time, as the bubble transgresses towards the upper wall. The area finally attains a zero value at t = 46185, indicating complete collapse of the first vapor bubble. The same cycle is repeated for each of the subsequent ones, allowing the domain to attain near-periodicity.

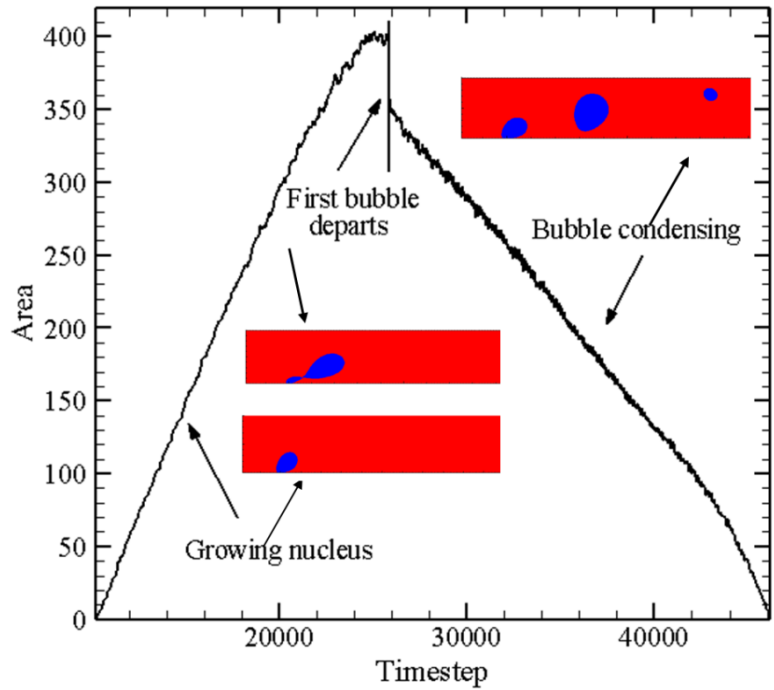

**FIG. 8.** (Color online) Temporal variation in the area of the first bubble over its entire lifespan, along with certain important snapshots

It may be noted here that the evolution of the shape of the first bubble till departure is visibly similar to the one reported by Sun, Li, and Yang[59] employing LBM for slowlyflowing fluid, where they tracked the interface with an extended Cahn-Hilliard equation. Their approach requires the assumption of an initial nucleus, while being computationally expensive. The infirmity of their model is highlighted by the regression relation of $D_d \propto g^{-0.425}$ for pool boiling. To expedite a better qualitative comparison, we have compared the bubble shapes during nucleation and growth with the observations of Zu et al.[28] in Fig. 9. They experimented with a rectangular microchannel of 0.38 mm height, housing a localized heater at the upper wall, and also performed numerical simulation adopting volume-of-fluid with pseudo-boiling approach, where vapor is injected into the domain at a controlled rate to simulate phase-change. The similarity in the shapes from nucleation till departure is quite palpable. A quantitative comparison, however, is not feasible, as their working medium was water, involving noticeable difference in working condition. Still the success of the present model in reproducing the physics around the nucleation site cannot be mistaken. A measure of the substantially greater computational resource requirement with volume-of-fluid solver is evident from the employment of $225 \times 50 \times 20$ meshes, with local grid-refinement around the point of injection, despite which it fails to replicate the natural process of nucleation.

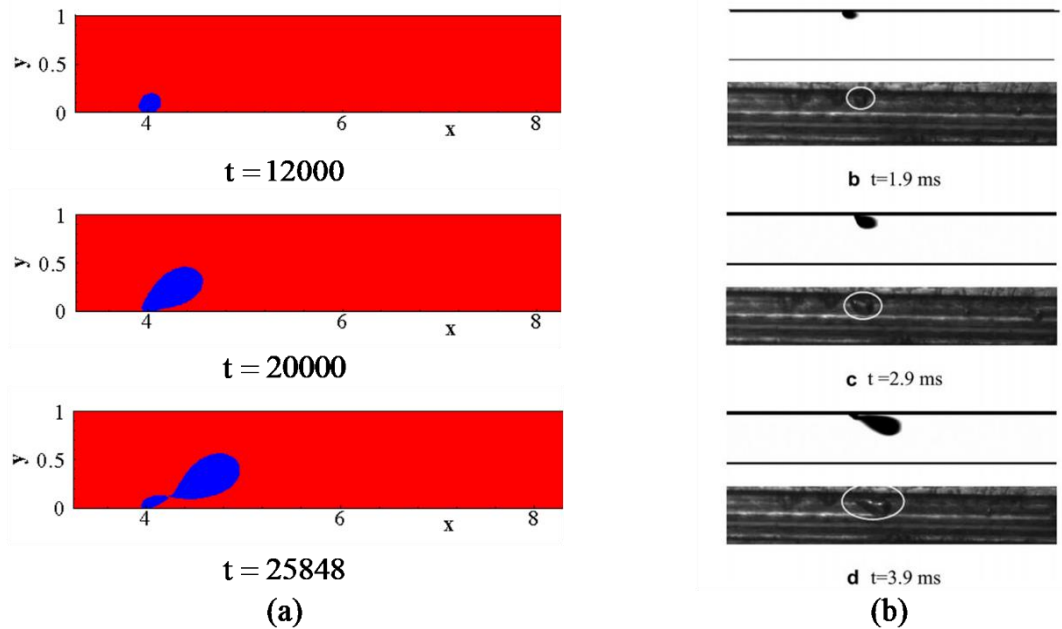

**FIG. 9.** (Color online) Visual comparison of the evolution in bubble shapes till departure with the experimental observations and computational results of Zu et al.[28]

The departure diameter and time required for departure are expected to hinge upon the prevailing drag force, which, in turn, is a direct function of the inlet velocity. The liquid inertia is higher with larger mass flux, and hence has greater potential of tearing the bubble from the microheater surface. Consequently, the departure diameter is supposed to reduce with increasing the flow rate, i.e., inlet Reynolds number. The converse is true for lower $Re_{in}$. Similar effect can be foreseen by increasing the wall superheat, as a larger surface temperature is apprehended to infuse energy at a faster rate into the nucleus, instigating accelerated bubble growth, and hence early departure owing to buoyancy. When the bubble diameter becomes comparable with the channel dimension, we enter the slug flow regime, characterized by large vapor slugs covering virtually the entire channel cross-section. The effect can be envisaged to be more prominent in narrow channels, as is the present case.

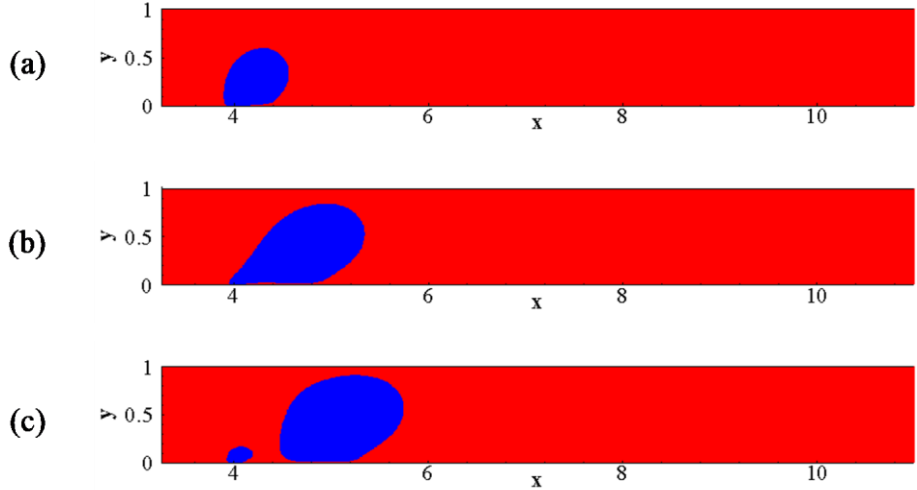

**FIG. 10.** (Color online) Snapshots of bubble evolution in slug flow ($Re_{in}$= 0.48), with red and blue colors respectively symbolizing liquid and vapor phases; (a) t = 18000: bubble growing on the heater surface at a fast rate; (b) t = 26000: bubble size comparable with the channel dimension and assuming a more elongated shape; (c) t = 30000: departure of an elongated bubble slug, which initially slides along the bottom wall, maintaining a small liquid film adjacent to the top wall.

In an effort to reproduce the above physics, simulations are performed for $Re_{in} = 0.48$, while maintaining identical superheat and contact angle, and a few sample snapshots of the domain is presented in Fig. 10. The growth of the bubble is much faster compared to the bubbly flow situation, as is shown in Fig. 10(a) for t = 18000. The growth rate continually increases with time, and the bubble starts elongating in the flow direction, as it has no space to expand in vertical. A notably-stretched bubble is about to depart at t = 26000 (Fig. 10(b)). Lower velocity results in reduced drag force, which delays the departure, while allowing the bubble to grow, emphasizing the role of liquid inertia. At t = 30000, departure is complete and the large vapor slug is sliding along the lower wall, while a small nucleus is left behind to initiate a repetitive pattern. A direct comparison can be drawn here regarding the transient evolution of the shape of the nucleus till departure for the two Reynolds number considered, which is available in Fig. 11. For $Re_{in} = 0.96$, the bubble starts to tilt towards the right almost immediately after appearance and gradually assumes an elliptical profile, while also shifting upwards owing to enhancing buoyancy effect. For $Re_{in} = 0.48$, however, the bottom interface of the nucleus continues to be very close to the lower wall, and the bubble slides along the wall even after the departure. Similar sliding of slug bubbles have been observed in microchannel during experiments as well[60]. Therefore, present SC-LBM algorithm is successful in capturing the multiphase dynamics inside the narrow channel for both the bubbly and slug flow regimes, while adhering to the underlying physics. Now we can employ the in-house code for exploring the effect of relevant parameters of importance.

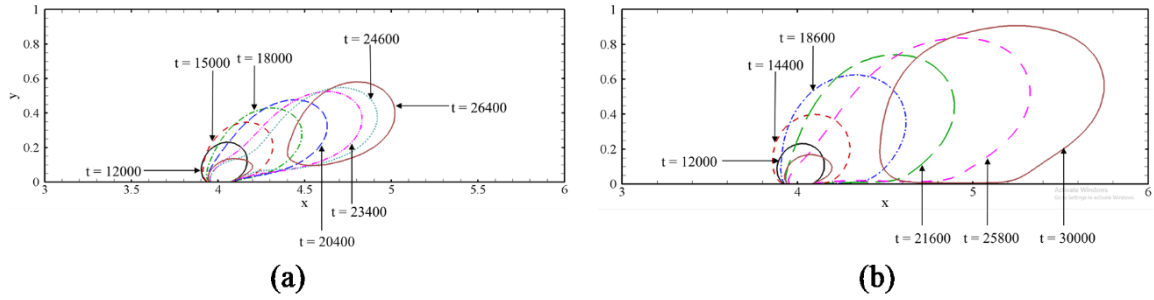

**FIG. 11.** (Color online) Temporal evaluation of the interface of the bubble for (a) $Re_{in}$ = 0.96 and (b) $Re_{in}$= 0.48

## *B. Effect of inlet velocity*

As discussed above, the supply velocity, characterized by the inlet Reynolds number $Re_{in}$, has a strong influence on the departure diameter and nucleation frequency. Greater velocity enhances the shearing effect, causing early departure with lesser bubble volume. Du et al.[58] argued that a larger flow rate also augments convective heat transfer, which allows a greater fraction of the supplied energy to be carried by the liquid phase, thereby slowing down the bubble growth. To have a more comprehensive view, simulations are performed for multiple $Re_{in}$, and the concerned variations in the average values of the departure diameter and time are presented in Fig. 12(a). Since the first bubble leaves the heater surface, the exercise becomes near-periodic, with minor variations in the departing bubble volume. Here we have estimated the departure diameter as the one averaged over five subsequent departures, while the time period is averaged over four successive releases since the first one. Both the departure diameter and ebullition time of a single bubble continually reduces with $Re_{in}$, in anticipated line. Yoo, Estrada-Perez, and Hassan[60] nicely deliberated about the conflict between the single-phase convection and bubble-induced heat transfer on increasing the mass flux. An enhanced flow rate affects both the wall evaporation and quenching processes, causing lesser drop in the temperature of the superheated liquid layer, allowing a quicker recovery towards subsequent nucleation. That can be viewed as the primary reason of the reduction in ebullition time with rise in flow rate. Sun, Li, and Yang[59] found the departure diameter to show an exponential relationship with the inlet velocity, where the release frequency varied linearly. Du, Zhao, and Bo[58], on the contrary, proposed a regression relationship of $D_d \propto Re_b^{-0.751}$, analyzing 5 different databases. Here, $D_d$ is the average departure diameter normalized using the characteristics length, and $Re_b$ is the bubble Reynolds number. We can relate the simulated departure diameters as, $D_d^{avg} \propto Re_{in}^{-0.635}$, which is reasonably close to Du, Zhao, and Bo[58], while the time period exhibits a near exponential decline.

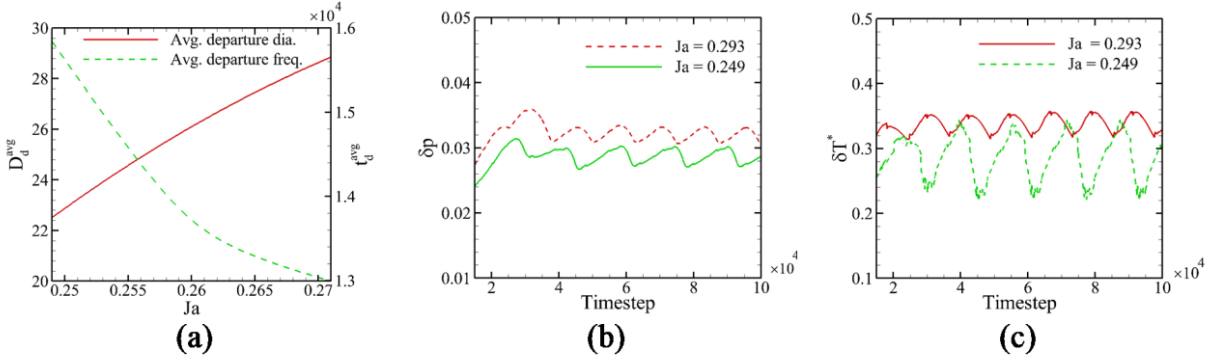

**FIG. 12.** (Color online) (a) The variations in average bubble departure diameter and average bubble departure time with inlet Reynolds number ($Re_{in}$), and temporal profiles of (b) inlet-to-outlet pressure differential and (c) normalized temperature difference for two different $Re_{in}$

The pressure difference across the channel also acquires a periodic pattern, as is displayed in Fig. 12(b), where the temporal variation in pressure drop along the centerline is available for two different $Re_{in}$. As mentioned earlier, emergence of the vapor phase enhances the pressure differential, with the accelerational component coming to prominence. It can clearly be seen that every departure corresponds to a reduction in pressure drop, whereas the same starts increasing again with the growth of the next nucleus. With the bubble occupying increasingly larger fraction of the flow area, the pressure drop increases, as the fluid density averaged over the entire domain starts to drop. Post-departure, the bubble condenses in the free stream, which helps the domain to recover pressure for a certain period. Such steep fall in pressure differential was also predicted by the 3D model of Zu et al.[28]. With increase in flow velocity, the shearing loss at both the walls gets augmented, incurring greater pressure loss, which can also be seen from Fig. 12(b).

The associated heat transfer rate can also be characterized following the temporal variation in the normalized temperature differential at the heated surface, $\delta T^* = (T_{wall} - T_{neighbour})/(T_{wall} - T_{bulk})$ as presented in Fig. 12(c). At the onset of nucleation, vapor starts accumulating on the heated surface, not allowing liquid to come in contact.

Vapor being a poor conductor, rate of heat transfer starts declining, as is manifested by the steep rise in the temperature differential. That is also one of the reasons of having stiff temperature gradient inside the nucleus itself (Fig. 7). Thermal communication starts to improve with the initiation of necking. As the bubble gets detached, surrounding liquid rushes in to fill that void, which strongly enhances the convective component of overall heat transfer. The same is manifested in the form of the rapid deterioration in $\delta T^*$. For lower $Re_{in}$, forced convection is weaker, reducing the heat transfer rate and thereby increasing the surface temperature differential (Fig. 12(c)). The slower bubble growth is also responsible, as vapor stays longer in contact with the heated surface. The upsurge in heat transfer is much starker at lower flow rates, owing to the release of larger-sized bubbles and consequently higher level of liquid motion. Similar role of Re on the heat transfer coefficient during flow boiling is well-documented in literature[60].

## C. Effect of wall superheat

The degree of superheat imposed on the microheater can be represented in terms of the Jacob number $Ja\left(=c_{p,f}\left(T_{wall}-T_{sat}\right)/h_{fg}\right)$, which can be viewed as the ratio of sensible heat transferred to the liquid at the wall to the latent heat transfer. All thermophysical properties being constant, Ja here is a direct function of the wall superheat. The effect of Ja on the flow field can be envisaged following Fig. 13. Du, Zhao, and Bo[58] mentioned that a higher Ja strengthens the bubble growth force, which resists the departure. Higher wall superheat also allows the nucleus to gain energy at an elevated rate. Both the above factors contribute towards greater departure diameter, as is evident from Fig. 13(a). A quicker growth also raises the buoyancy, which aids the release from the nucleation site, consequently reducing the length of the ebullition period. The temporal profiles of pressure drop across the channel and normalized temperature difference follows the periodic behavior, as discussed above. Higher wall superheat corresponds to greater bubble area and also enhanced bubble-induced liquid motion post-departure, which is reflected in elevated inlet-to-exit pressure differential. It is interesting to observe that, on increasing Ja, the highest level of $\delta T^*$ remains nearly the same. However, the undulations are substantially subsided, yielding more-uniform heat transfer rate over the entire span under consideration. That is possibly a consequence of the quicker bubble release and enhanced bubble induced heat transfer[60,61].

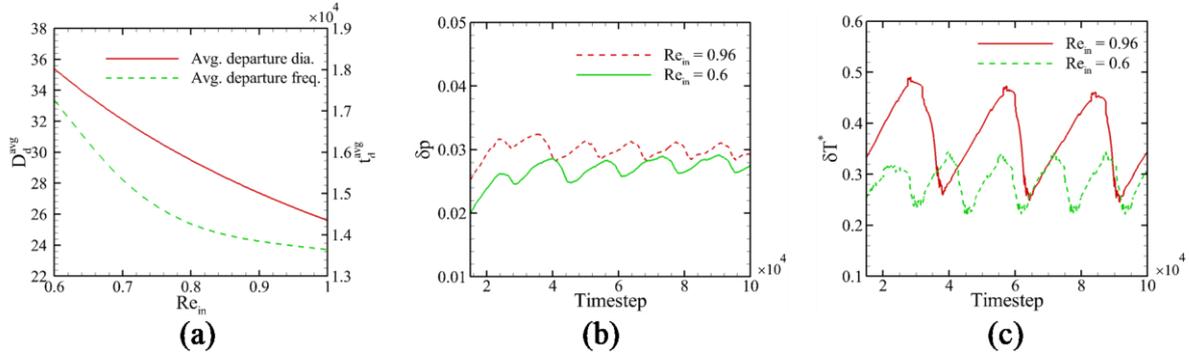

**FIG. 13.** (Color online) (a) The variations in average bubble departure diameter and average bubble departure time with Jacob number (Ja), and temporal profiles of (b) inlet-to-outlet pressure differential and (c) normalized temperature difference for two different Ja

## D. Flow Regime Map

We have seen from the previous discussions that the present SC-LBM algorithm is successful in reproducing the two most common multiphase flow regimes associated with horizontal narrow channels, and also logical pattern of parametric effects. The combined effect of the two variables considered here, namely $Re_{in}$ and Ja, can be summarily viewed in terms of a flow regime map. We have already observed that the bubble departure diameter continually increases with reduction in inlet Reynolds number, owing to the reduced shearing action, and it is possible to attain the slug flow condition for sufficiently small $Re_{in}$. Similarly, a greater Jacob number produces larger bubble at the instant of release, as higher rate of energy addition infuses more energy into the nucleus. The same is distinctly demonstrated in Fig. 14. If the flow rate is very high or degree of superheat is low enough, sufficient amount of energy is not available to induce the phase change, and the entire domain experiences solely single-phase flow. Consequently, moving along the negative x-direction or positive y-direction, we can sequentially encounter single-phase flow, multiphase bubbly flow and multiphase slug flow respectively. Same can also be achieved following the arrow shown in the figure, which indicates the direction towards which the departure diameter increases. Qualitatively similar role of inlet mass flux regarding transition from bubbly to slug flow was hinted by Harirchian and Garimella[62] for horizontal microchannels of various dimensions during their experiments.

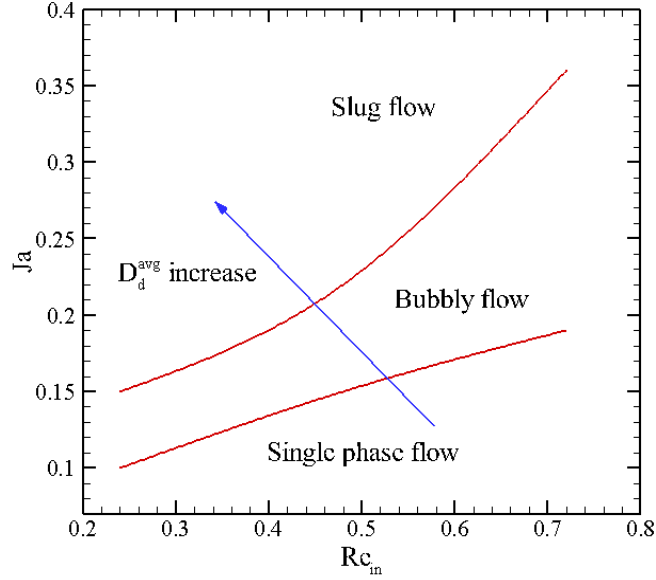

**FIG. 14.** (Color online) Flow pattern map in $Re_{in}$ - $Ja$ plane; where the flow regime changes from single-phase to bubbly to slug in the direction of the arrow; the average bubbled departure diameter increase with increase in $Ja$ and decrease in $Re_{in}$

## *E. Effects of surface wettabiltiy*

For a specific fluid flowing under a particular set of operating conditions, the nature of the three-phase interface can be a deciding factor in determining the ebullition characteristics and consequent channel thermohydrodynamics. Any combination of liquid-vapor interface and solid surface has an equilibrium value for the static contact angle at given pressure and temperature, dictated by the interfacial energy balance, which characterizes the wettability of the surface. As already menioned through Eq. 15, the contact angle can be modulated here by varying the false wall density $\rho_w$, a graphical representation of which is available in Fig 15. For the present PREOS at $0.9T_c$, the lattice-level densities of the liquid and vapor phases are 5.95 and 0.59 respectively. With $\rho_w$ assuming a value closer to the liquid one, the contact angle approaches zero, whereas the reverse is true when it moves towards vapor level.

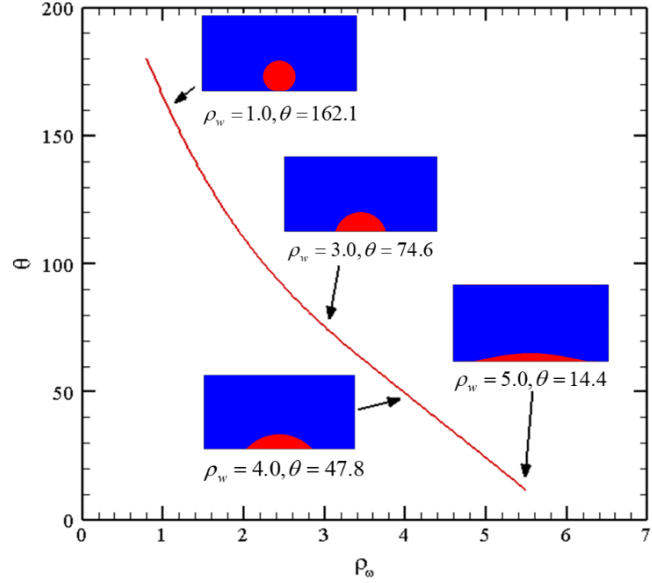

**FIG. 15**. (Color online) Variation of the static contact angle $\theta_c$ with the false wall density $(\rho_w)$, where some sample bubble shapes are also shown; an increment in $\rho_w$ causes a decrement in $\theta_c$

Mukherjee et al.[61] argued that the bubble departure is determined by the instantaneous drag force of the passing liquid and the prevailing surface tension force at the bubble contact line. A larger contact angle corresponds to lesser surface wetting by the liquid phase, allowing better accumulation of vapor on the microheater. With vapor covering larger portion of the heated surface, there is an increase in the effective heat transfer area available to the nucleus[63]. It is, therefore, feasible for the bubble to absorb larger amount of energy, leading to higher departure diameter. The same can be confirmed from Fig. 16(a-b). A reduction in the contact angle $(\theta_c)$ from $64.74^0$ to $52.23^0$ inflicts a noticeable change to the shape at the instant of departure, as well as hints towards a possible shift in the flow regime altogether. While $\theta_c = 52.23^0$ yields smaller bubble, leading to the bubbly flow pattern, the other one produces an elongated bubble, which tends to slide along the surface initially, and then approaching the slug flow regime. The effect of the contact angle on the departure diameter is summarized in Fig. 16(c). $D_d$ increases almost linearly with $\theta_c$ and an augmentation of about $17^0$ enforces more than 12% rise. An increase in heat transfer area also enhances the bubble growth force, which tends to resist the departure. Consequently the average bubble release period steadily increases with $\theta_c$.

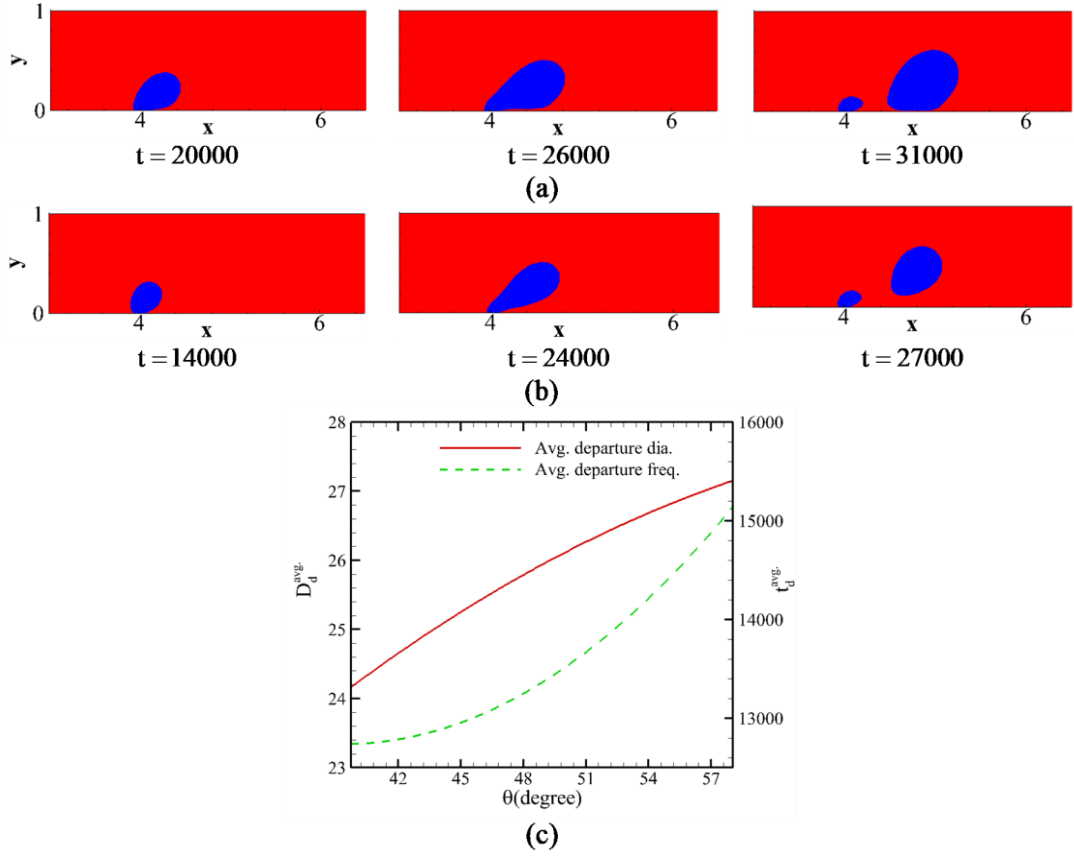

**FIG. 16.** (Color online) Effect of surface wettability on the bubble characteristics at the instant of departure. The snapshots of bubble growth till departure for (a) $\theta_c = 64.74^0$ and (b) $\theta_c = 52.23^0$, clearly shows that the departure diameter increases with the contact angle. (c) The variations in average bubble departure diameter and average bubble departure time with the contact angle

Periodic dewetting and rewetting of the heated surface is another common phenomenon in conjunction with flow boiling. During the bubble growth, liquid phase converts to vapor at the microheater surface, pushing liquid away from it, which is referred as dewetting. On initiation of necking, however, liquid is allowed to rush back towards the middle of the heater, thereby rewetting it again, which leads the path towards the next nucleation. This repetitive pattern with the three-phase contact line is clearly visible from Fig. 17, in terms of the temporal evolution of its length $(L_t)$. As already explained, vapor phase covers larger area with the increase in the contact angle, causing an obvious increase in the length of the three-phase contact line. $L_t$ increases from a peak value of about 6.8 for $\theta_c = 52.23^0$ to close to 10 for $\theta_c = 64.74^0$. The recurrent nature can be correlated using a regression relation of the form, $L_t = L_0 + a \sin\left(\pi \frac{t - t_c}{w}\right)$, where $L_0, a, t_c, w$ are all functions of

$\theta_c$. For example, with $\theta_c = 52.23^0$, we have $L_0 = 5.65$, $a = 1.0276$, $t_c = 4398.68$, and $w = 6844.05$. We can, therefore, claim that the present algorithm is capable of efficiently capturing the effect of surface wettability as well.

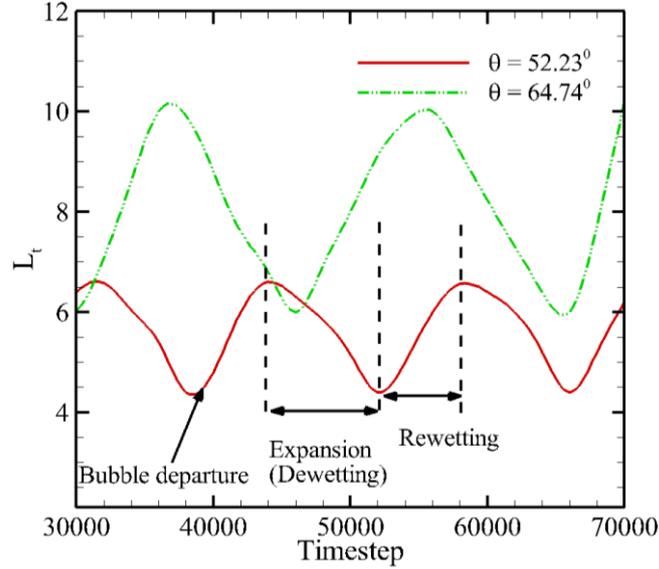

**FIG. 17.** (Color online) Effect of the contact angle on the movement of the three-phase contact line of the growing bubble

## 4. Conclusions

Precise numerical estimation of flow boiling has been a concern for the multiphase community over the years, despite significant development in resources. Conventional Eulerianaveraging-based macroscopic approach generally requires an assumed form of the interface, making them unsuitable for emulating complex boiling scenarios. The mesoscopic latticeBoltzmann method, adopting Boltzmann statistical averaging, provides a potent alternate option, and has gathered substantial attention in recent years. Its application, however, has mostly been limited to pool boiling, with only sporadic attempts to replicate flow boiling. Present study aims towards filling that void following a pseudopotential-based LB approach for computational appraisal of a 2D narrow fluidic channel accommodating a microheater on the lower wall. Adopted boundary conditions of fixed outlet pressure and specified inlet temperature combines to inflict the characteristics of subcooled flow boiling and the present SC-LBM algorithm successfully reproduces the qualitative attributes of the same.

The entire thermohydrodynamics associated with flow boiling in narrow channels, involving nucleation, growth and departure of the bubble at the microheater, and subsequent adoption of bubbly or slug flow regimes, have been aptly illustrated. Both the departure diameter and ebullition period of a bubble at the nucleation site are dependent on the inlet velocity and wall superheat. Present algorithm is able to capture their effects quite logically, while also being consistent with the available literature. We have compared the bubbly and slug flow regimes in terms of the bubble growth pattern and departure characteristics. Pressure drop across the channel and heat transport characteristics exhibit a repetitive pattern, which is also a rational observation. Increase in wall superheat or reduction in mass flux tends to convert the flow domain from single-phase to bubbly to slug, which is amply demonstrated through a qualitative flow regime map. Finally the effect of surface wettability in terms of the contact angle has also been discussed, and again the SC-LBM algorithm is able to yield logical prognosis. So we can conclude that the pseudopotential-based LBM has sufficient competency of simulating flow boiling scenarios and is expected to find applications in more complicated domain.

One of the apprehensions with the SC-LBM algorithm is possible thermodynamic inconsistency involving large density ratios, which limits its employment primarily to large values of the reduced temperatures, such as $0.9T_c$ adopted for our work. Addressing the same can be viewed as the next logical step of the present study, while the configuration can also be extended to involve multiple nucleation sites or a distributed heater.

**Acknowledgements**

All the computations reported here are carried out in the PARAM-ISHAN cluster, a 162 nodes, 250 tfps hybrid high performance computing facility at IIT Guwahati. PKM also acknowledges the financial support provided by the SERB (DST), India, through project No. ECR/2016/000702/ES.